# The Global Surface Roughness of 25143 Itokawa

Hannah C. M. Susorney[a], Catherine L. Johnson[a], Olivier S. Barnouin[b,c], Michael G. Daly[d], Jeffrey A. Seabrook[d], Edward B. Bierhaus[e], Dante S. Lauretta[f]

[a]*Department of Earth, Ocean and Atmospheric Sciences, University of British Columbia, Vancouver, BC V6T 1Z4, Canada.*
[b]*The Johns Hopkins University Applied Physics Laboratory, Laurel, MD 20723, USA.*
[c]*Hopkins Extreme Material Institute, The Johns Hopkins University, Baltimore, MD 21218, USA.*
[d]*The Centre for Research in Earth and Space Science, York University, Toronto ON M3J 1P3, Canada.*
[e]*Lockheed Martin Space Systems Company, Denver, CO., USA.*
[f]*Lunar and Planetary Laboratory, University of Arizona, Tucson, AZ., USA.*

## Abstract

Surface roughness is an important metric in understanding how the geologic history of an asteroid affects its small-scale topography and it provides an additional means to quantitatively compare one asteroid with another. In this study, we report the first detailed global surface roughness maps of 25143 Itokawa at horizontal scales from 8–32 m. Comparison of the spatial distribution of the surface roughness of Itokawa with 433 Eros, the other asteroid for which this kind of analysis has been possible, indicates that the two asteroids are dominated by different geologic processes. On Itokawa, the surface roughness reflects the results of down-slope activity that moves fine grained material into geopotential lows and leaves large blocks in geopotential highs. On 433 Eros, the surface roughness is controlled by geologically-recent large impact craters. In addition, large longitudinal spatial variations of surface roughness could impact the role of YORP on Itokawa.

---

## 1. Introduction

Asteroids are a diverse set of solar system small bodies that have been investigated using both spacecraft and Earth-based observational platforms. To better understand processes that shape the surface histories of these bodies, we seek quantitative metrics to investigate the geology of individual asteroids and to compare one asteroid to another. One such metric is surface roughness, which quantifies the topography of asteroids using a measure of the change in topography over a specified horizontal scale. Surface roughness thus provides a statistical measure of topographic variability.

The surface roughness of asteroids can be obtained over different spatial scales, known as baselines, using a variety of measurement techniques. Surface roughness is calculated at sub-centimeter to centimeter scale from thermophysical data (e.g., Rozitis and Green, 2012), and is also derived from radar data at the scale of centimeters (e.g. Benner et al., 2008). Thermophysical and radar-derived surface roughness measurements are usually made with Earth- and space-based observational platforms.

At larger baselines, typically meters to tens of meters, surface roughness is calculated from spot-to-spot topographic variations from the highest resolution topographic data available, usually from image-based shape models, or laser altimeter measurements. Laser altimetry-derived measurements of surface roughness are limited to a few asteroids, such as 433 Eros and 25143 Itokawa, hereafter Eros and Itokawa respectively,(Cheng et al., 2002; Barnouin-Jha et al., 2008b). By comparing longer baseline spacecraft-derived surface roughness measurements with shorter baseline Earth-based-observation-derived surface roughness, we may be able to establish links between Earth-based observations and the attributes of the in-situ geology of an asteroid that might be later observed by spacecraft. In this study, we focus on laser altimetry derived surface roughness for Itokawa. The maps presented here are the first detailed meter-scale surface roughness maps of Itokawa, and in fact the first such maps for a small (sub-km) asteroid.

The Hayabusa spacecraft explored the asteroid Itokawa from September to December 2005 (Fujiwara et al., 2006). Itokawa is a near-Earth S-class asteroid (Binzel et al., 2001), which is 535 × 294 × 209 m in size (Demura et al., 2006). Its shape is often described as bilobate, with a distinct head and body (Abe et al., 2006). On the basis of the estimated bulk density of 1.9 g/cm$^3$, Itokawa is interpreted to be a rubble-pile (Fujiwara et al., 2006), with an interior that is likely the re-accumulated debris of a past catastrophic impact (Fujiwara et al., 2006). Itokawa (Fig. 1) can be broadly split into two regions, the highlands (Fujiwara et al., 2006; Abe et al., 2006; Saito et al., 2006; Demura et al., 2006; Yano et al., 2006; Miyamoto et al., 2007) and lowlands (the Muses–Sea and Sagamihara regions; Fujiwara et al., 2006; Saito et al., 2006; Demura et al., 2006). The highlands are covered in blocks ranging in size from 10s of centimeters to the largest boulder Yoshinodai, which is 50 × 30 × 20 m (Saito et al., 2006). The lowlands are covered in grains that are centimeter- to millimeter-sized, inferred from the Hayabusa spacecraft touchdown location in the Muses–Sea (Yano et al., 2006) and return samples (Nakamura et al., 2011). The distribution of blocks follows the geopotential of the surface of Itokawa with small grains concentrated in the lowlands and the larger blocks in the highlands (Fujiwara et al., 2006; Saito et al., 2006; Miyamoto et al., 2007; Noviello, J L et al., 2014; Tancredi et al., 2015). Impact craters have been identified on the surface, although they are often difficult to distinguish from the surrounding terrain and have a shallower depth/diameter ratio than on other asteroids (Hirata et al., 2009). This may reflect either different inherent crater morphology or modification due to movement of, and filling by, fine-grained regolith material.

The surface roughness of Itokawa was first explored using radar circular polarization ratios by Ostro et al. (2004) who found similar radar-scale-surface roughness properties (approximately 13 cm baseline) to those of Eros, a 30 × 10 × 10 km asteroid explored by the Near Earth Asteroid Rendezvous (NEAR)-Shoemaker mission (Zuber et al., 2000). Onboard the Hayabusa spacecraft was a LIght Detection and Ranging (LIDAR) instrument that collected detailed topography data from the surface of Itokawa (Abe et al., 2006). Initial local surface roughness studies using a few regional LIDAR tracks found that the highlands and lowlands had larger and lower, respectively, surface roughness values than those of Eros at similar baselines (Abe et al., 2006; Barnouin-Jha et al., 2008b). Barnouin-Jha et al. (2008b) also calculated the thickness of the mobile regolith of Itokawa in the Muses–Sea (2.3 ± 0.4 m) from the difference in surface roughness of the highlands and lowlands. The volume of regolith estimated from this result corresponds to an equivalent global layer ∼0.4 m thick. A preliminary global investigation of surface roughness across Itokawa (Barnouin-Jha et al., 2008a) (mapped in spatial bins of 15 by 15 degrees) confirmed these results. More recently, (Muller et al., 2014) found that the observed thermal properties cannot be explained by the current Itokawa shape model, and that higher-resolution centimeter-scale surface roughness must also be present. This higher-resolution centimeter-scale surface roughness is likely present in the lowlands which are dominated by centimeter- to millimeter-sized particles (Yano et al., 2006). High resolution images of large boulders on Itokawa indicate that the surface of some boulders have a rough undulating texture (Noguchi et al., 2010).

In this study, we conduct the first complete detailed global assessment of meter-scale surface roughness on Itokawa. We generate the first global surface roughness dataset for Itokawa using the entire LIDAR dataset returned by the Hayabusa mission (Barnouin-Jha et al., 2008b). We investigate whether and how the surface roughness of a rubble-pile differs from the previously measured global surface roughness of Eros, a fractured monolith (Susorney and Barnouin, 2018). Additionally, we discuss whether the surface roughness at meter-scales might be related to centimeter-scale surface roughness; both the meter and centimeter scale is important when considering YORP effects that can alter the spin state and shape of asteroids (e.g., Rozitis and Green, 2012). The results of this study will also allow the surface roughness of Itokawa to be quantitatively compared to that of other small asteroids currently being explored, such as 162173 Ryugu (Tsuda et al., 2013) and 101955 Bennu (Lauretta et al., 2015).

In the following sections, we summarize how we adapt the methodology for calculating surface roughness on asteroids (Cheng et al., 2002; Barnouin-Jha et al., 2008b; Susorney and Barnouin, 2018) to the global Hayabusa

LIDAR dataset (Section 2), to produce the first detailed global surface roughness maps for Itokawa. We present the global maps and surface roughness results (Section 3), and discuss their implications for the geology of Itokawa and for future directions in asteroid surface roughness research (Section 4). Finally, we discuss the major conclusions of this study (Section 5). In this paper we refer to the lowlands of Itokawa as the Muses–Sea region for consistency with previous studies (e.g., Barnouin-Jha et al., 2008b), however we note that the IAU registered name is MUSES-C Regio.

## 2. Methodology

In this study, we use root-mean-square (RMS) deviation (Shepard et al., 2001) as our measure of surface roughness for direct comparison with the results of a previous localized study of Itokawa (Barnouin-Jha et al., 2008a), as well as global and regional studies of Eros (Cheng et al., 2002; Susorney and Barnouin, 2018). RMS deviation is commonly used to assess the influence of geological processes on surface roughness for larger planetary bodies such as Mercury and the Moon (e.g., Kreslavsky et al., 2014; Rosenburg et al., 2011), providing useful data for comparisons with, and interpretation of processes acting on, asteroids such as Itokawa. We also use RMS deviation because it is directly related to the commonly reported measure of surface roughness, namely RMS slope (see Shepard et al., 2001, for the relationship), which is used in thermophysical models. Additionally when a surface is found to be self-affine, the Hurst exponent derived from RMS deviation (see below), allows measurements of RMS deviation obtained at one set of horizontal baseline scales to be used to estimate the RMS deviation at other baselines.

This, in turn, allows comparison of surface roughness measurements across an individual body, as well as comparison of measurements among different bodies.

RMS deviation, $v(L)$, is defined as the change in detrended topographic height, $h$, over a given horizontal baseline, $L$, and is defined by

$$v(L) = \left[\frac{1}{n}\sum_{i=1}^{n}[\Delta h(L)_i]^2\right]^{\frac{1}{2}}, \qquad (1)$$

where $\Delta h(L)_i$ is the $i$'th estimate of a change in height and $n$ is the number of $\Delta h$ used to calculate RMS deviation. RMS deviation can be calculated on a point-wise basis to produce maps of spatial variation in surface roughness or averaged regionally or globally. RMS deviation is related to the Hurst exponent, $H$, which describes how the surface roughness changes with increasing baseline by

$$v(L) = v_o L^H, \qquad (2)$$

where $v_o$ is the RMS deviation at the unit scale. The RMS deviation metric is unstable at small $n$ (Kreslavsky et al., 2013; Susorney et al., 2017; Susorney and Barnouin, 2018) and it is necessary to quantify how large $n$ in equation (1) must be to obtain a stable estimate of $v(L)$. We find that for $n \geq 200$, the estimate of $v(L)$ converges (Fig. 2), indicating stability.

*2.1. Calculating surface roughness from the Hayabusa LIDAR*

Individual LIDAR tracks were used rather than a global gridded topographic dataset (e.g., derived from a shape model) for computing surface roughness because global gridded datasets for Itokawa included registration errors between different LIDAR tracks. Straight LIDAR tracks across Muses–Sea Regio indicate that the vertical resolution of the LIDAR instrument is 0.5 m (Barnouin-Jha et al., 2008b). We used the spatially registered LIDAR tracks available for Itokawa (Mukai et al., 2012). These have been corrected for uncertainties in spacecraft location and, as verified with images, are now registered at the correct locations on the asteroid (Barnouin-Jha et al., 2008b). The individual LIDAR footprints on Itokawa were calculated over a 8 m area (Barnouin-Jha et al., 2008b) therefore we chose 8 m as our minimum baseline.

To calculate surface roughness we first calculated the topography (also called geopotential altitude; Scheeres et al., 2016) using the method described in (Cheng et al., 2002; Barnouin-Jha et al., 2008b) for individual LIDAR returns using the 49,152-plate shape model of Itokawa that was derived from stereo-photoclinometry (SPC). We used the reported density of 1.95 g/cm$^3$, a rotation rate of 0.000144 rad/s, and a reference potential of -0.0147 J/kg (Abe et al., 2006), and a polyhedral method for calculating the gravitational potential at fixed points on an irregular body (Werner and Scheeres, 1997)

Due to the loss of several reaction wheels the Hayabusa spacecraft was unable to stabilize. The resulting LIDAR tracks appear to wander and twist across the surface of Itokawa (Fig. 3). This makes it difficult to develop an accurate metric required to define the distances between topographic heights, and thus the baseline, an essential ingredient for assessing surface roughness. Previous studies that calculated the surface roughness of asteroids used only the most straight altimetric tracks, together with the spacecraft clock, to approximate distance across the asteroid, e.g. for Eros (Susorney and Barnouin, 2018) and Itokawa (Barnouin-Jha et al., 2008a). We therefore explored several methods to approximate distance for the global Itokawa dataset.

To calculate the surface roughness on a track-by-track basis, first, we found all the points within 10 baselines (the distance, calculated as a simple Euclidean distance) from the point of interest (a single LIDAR return) along the track. We refer to this region as the region of interest. The distance between the points in this region and the point of interest was calculated using four different distance metrics ('straight-along-track','utc-straight-along-track','radial-distance', and 'plane-radial', Fig. 4). In the 'straight-along-track' distance metric we measured the xyz distance from LIDAR point to LIDAR point along-track. In the 'utc-straight-along-track' distance metric we projected the LIDAR points onto a straight line and used the spacecraft clock of each LIDAR pulse to estimate distance along this straight line (used for Eros by Susorney and Barnouin (2018)). Neither of these two distance metrics work well for the Hayabusa dataset because they either elongated ('straight-along-track') or compressed ('utc-straight-along-track') the distance between points. The 'radial-distance' metric used the Euclidean distance from the point of interest to each of the other points in our region of interest. In the 'radial-plane' metric we projected the LIDAR points in the region of interest onto a plane that was best fit to the LIDAR points from the track in the region of interest and calculated the Euclidean distance on the plane. We find that the radial-plane distance metric compressed the distance between LIDAR points in regions of high curvature, such as the ends of the asteroid. Over larger distances the radial-distance metric also compressed the distance between points, but to a smaller extent than the radial-plane method, but this was not an issue at smaller baselines.

We compared maps of all four of our distance metrics and found that spatial roughness variations among the metrics (Fig. S1-S6) differ substantially (greater than 5 m at $L = 32$ m) at baselines greater than 32 m and larger, thus we set 32 m as our maximum baseline. We chose to use the radial-distance metric as our distance metric, because as described above it was less affected by artificial contraction or elongation than the other metrics and it is straightforward to understand. However, because the absolute surface roughness values depends on the distance metric (Fig. S1-S6), we are conservative in quantitative interpretation of the absolute surface roughness values derived.

After calculating the distance between LIDAR points in our region of interest using the radial-distance metric, we fitted a plane to the points and rotated the points to a new coordinate system in which the z-axis is parallel to the normal to the plane. The transformed x, y, and topography values were then used to detrend the topography by fitting a plane to x, y, and topography and rotating the plane and points such that the normal to the plane was along the z-axis, i.e. the plane's x and y values equaled zero. Finally, we found the point closest to one baseline along the track (in a forward sense) from the original point of interest, and measured the difference in height, $\Delta h$. We repeated this analysis for all LiDAR points on all tracks.

*2.2. Mapping RMS deviation*

The surface roughness of Itokawa was visualized by mapping it onto the SPC shape model that we re-sampled down to a 10,000 plate model. To do this, we found all $\Delta h$ two baselines away from the center of each plate in the 10,000 plate model and calculated the RMS deviation (from Eqn. (1)) using the $\Delta h$ within two baselines from the center of each plate. We also checked that at least 200 $\Delta h$ were present on each plate to ensure a stable estimate of RMS deviation (Fig. 2). Changing the resolution of the shape model did not affect the surface roughness maps (Fig. S7).

## 3. Results

Global surface roughness maps of Itokawa were calculated at 8, 16, 24, and 32 m baselines. The 8 m and 32 m baseline maps are presented here for the radial-distance distance metric. The lower limit of 8 m was set by the averaging inherent in the LIDAR footprint, the upper limit of 32 m guided by the results discussed above. The results for the intermediate baselines of 16 m and 24 m are not shown, but were examined visually for consistency with the results reported here and were used in the calculations of Hurst exponents.

At the smallest baseline mapped (8m, Fig. 5), we observe a bimodal variation in surface roughness with the lowlands having lower surface roughness values relative to the highlands. Surface roughness is elevated at large boulders such as Yoshinodai. At the 32 m baseline (Fig. 6), we observe a similar distribution of surface roughness to the 8 m baseline with the Muses–Sea having lower surface roughness than the highlands on the head. The relative difference in surface roughness between larger boulders and the surrounding region is smaller than that at the 8 m baseline.

We used the surface roughness values at the four baselines at each plate on the 10,000-plate shape model to estimate a local Hurst exponent. The resulting map shows a large range in Hurst exponents across the surface (Fig. 7) from values near 0 (roughness invariant with baseline) to 1.0 (roughness self-similar at all baselines). The global Hurst exponent for Itokawa, calculated from all $\Delta h$s in this study, is $0.51 \pm 0.07$ and $v_o$ (equation 2) is $0.269 \pm 0.075$. This is lower than the Hurst exponent calculated for Eros, $0.97 \pm 0.01$ (Susorney and Barnouin, 2018). The lower Hurst exponent means that the ratio of surface roughness at the largest versus the smallest baselines is lower on Itokawa than on Eros.

*3.1. Correlations with other geologic features and asteroid properties*

We investigated whether there are correlations between the surface roughness, geologic features, and surface properties to understand how surface roughness could be related to the geology of Itokawa. Correlations were calculated using a Pearson's correlation coefficient and the surface roughness and the geologic property/features mapped onto the same shape model used to calculate surface roughness. Statistical significance was taken as $p < 0.05$, where $p$ is the probability of obtaining a correlation coefficient as large as that observed or greater, if the underlying distributions have no correlation. We found statistically significant positive or negative correlations for all the quantities examined; however in two cases (see below) the correlation coefficients, while meeting the significance criterion were much lower than in all other cases, and thus we did not interpret these results. Correlation coefficients are given in Table 1 for the physical quantities examined. Using the block counts from Mazrouei et al. (2014), surface roughness was found to show a statistically significant, positive correlation with block spatial density at the 8 m and 16 m surface roughness baselines. For longer baselines, 24 m and 32 m, the correlation coefficients, although meeting the significance criterion were much lower ( 0.1), suggesting some, but very weak correlation of block counts and roughness at these baselines. We also calculated topography, slope, and gravitational acceleration from the 49,152 plate shape model of Itokawa with these values remapped onto the 10,000 plate shape to compare with our surface roughness maps. A statistically significant positive correlation was found between surface roughness and slope, and between surface roughness and topography at all baselines examined. Gravitational acceleration is negatively correlated with surface roughness at all baselines examined.

*3.2. Thickness of regolith*

Barnouin-Jha et al. (2008b) estimated the thickness of regolith in the lowlands on Itokawa to be 2.3 ± 0.4 m by comparing the surface roughness of the highlands and lowlands. This estimate was based on the assumption that variations in surface roughness at the smallest baselines were due to regolith motion across the surface. In the previous study, surface roughness was measured on individual tracks that were clearly in the highlands and lowlands. As this study is global in nature we did not specifically map surface roughness in the highlands versus the lowlands, as there is no currently-agreed-upon definition of a boundary separating these regions. We instead make a qualitative global estimate of the thickness of regolith in the lowlands using three different measures. First, the absolute range in RMS values for the 10,000 plate model at the baseline of 8 m results in an estimate of regolith thickness of 10.6 m and provides an upper bound of topography that is buried by mobile regolith in the lowlands. Second, the interquartile range (the range between the 25th and 75th percentile) for the same dataset is 0.8 m and provides a more conservative estimate of reguluth thickness. Third, visual estimates of surface roughness values in the highlands versus the lowlands differ by about 1 m. The values obtained encompass the estimates made by Barnouin-Jha et al. (2008b).

## 4. Discussion

The high-resolution surface roughness maps on Itokawa can be used to understand the surface geology of Itokawa and to compare Itokawa to other asteroids. While, there are issues with the LIDAR dataset (see section 2.1) that prevented distance from being measured in the same manner as on Eros, we can still compare different regions of Itokawa and make qualitative comparisons with the surface roughness of Eros.

### 4.1. Implications for the geology of Itokawa

The spatial distribution of surface roughness on Itokawa can be used to explore the role of different geologic processes in controlling topography at the horizontal scales investigated here. Previous studies of Itokawa (Barnouin-Jha et al., 2008b) found a dichotomy in surface roughness and proposed that this resulted from the downslope movement of smaller grains from the highlands to the lowlands, filling regions of lower topography. Our global maps of surface roughness from 8–32 m support this previous inference. We find that the surface roughness at the 8 and 16 m baselines are positively correlated with block spatial density and that the surface roughness at all baselines is correlated with slope and geopotential. Thus the surface roughness at the few meters to tens of meter scale on Itokawa is dominated by the current shape, and geopotential of the asteroid.

We used spatial variations in surface roughness to provide rough estimates of regolith thickness globally. Our results give a layer of mobile regolith on the order of 1 m, with an upper bound of ∼ 10 m. Thus both this work and that of Barnouin-Jha et al. (2008b) point to a layer of mobile regolith on Itokawa that is a meter to a few meters thick.

We also investigated whether there are any differences in the surface roughness of the head and body of Itokawa to see if we could identify any evidence that the head and body evolved separately or together (see the discussion in Mazrouei et al., 2014). No such differences were found, implying that if there is a difference in internal structure between these two regions, it does not affect the surface geology, which is instead dominated by the highlands/lowlands dichotomy. If the head and body of the asteroid evolved independently, any trace of differing topography at the scale of 8 − 32 m has been overprinted by down-slope motion that homogenize characteristics at these length scales. This is consistent with the correlation of surface roughness and geopotential implying that the current surface roughness of Itokawa is dominated by its current shape and the surface roughness is not affected by past states of Itokawa (i.e., past spin states or the parent body of Itokawa). The surface roughness of Itokawa reflects the current spatial distribution of boulders, which in turn are governed by geopotential highs and lows. If the shape or spin of Itokawa was different in the past, this would likely have resulted in a different distribution of boulders and thus surface roughness.

*4.2. Surface roughness and YORP on Itokawa*

As Itokawa is a relatively small asteroid it has been predicted that its spin can be affected by YORP (e.g. Scheeres et al., 2007). The modeling of the ex-pected YORP spin-down (Scheeres et al., 2007) is at odds with the observed YORP spin up (Lowry et al., 2014). This mismatch has been proposed to result from density variations between the head and body of Itokawa (Lowry et al., 2014), which could alter the YORP effect (Scheeres and Gaskell, 2008; Lowry et al., 2014). Although, the role of YORP on Itokawa is not fully understood (i.e., Ševeček et al., 2015), spatial variations in the surface roughness at the centimeter scale can alter the ability of YORP to modify the rotation rates of asteroids (Rozitis and Green, 2012). On Itokawa, we observe large longitudinal and latitudinal variations in surface roughness across Itokawa (Fig. 9). For example, at the 8 m baseline, a pole-to-equator increase in surface roughness of almost an order of magnitude and more localized longitudinal variations of a factor of ~2 are seen, (Fig. 9). Latitudinal variations in block density have also been noted on Itokawa, with enhanced areal boulder densities in the equatorial regions compared with at higher latitudes (Mazrouei et al., 2014). This is in part, reflected in the positive correlation we observe between surface roughness and boulder density. The longitudinal variations, while a smaller magnitude than latitudinal variations, are more likely to affect YORP (Rozitis and Green, 2012) and thus may contribute to the mismatch in observed and predicted YORP. The source of these spatial variations is likely variations in boulder densities within each longitudinal bin. If variations in surface roughness continue down to the centimeter scale, particularly in longitude, they will likely affect the thermophysical properties of Itokawa, and in turn, YORP.

*4.3. Comparison of Itokawa and Eros*

Comparisons of the surface roughness of Itokawa and Eros at similar baselines can provide constraints on differences in surface structure (and underlying processes) of a fractured monolith and a rubble-pile. The surface roughness of Itokawa is bimodal at baselines of 8–32 m, with the two modes associated with surface roughness of the highlands and the lowlands. In contrast, the surface roughness of Eros at baselines of 4–200 m does not show a clear differentiation by terrain type (Susorney and Barnouin, 2018). Although the surface roughness of Itokawa and Eros are both correlated with block spatial density at meter-scale baselines, the distribution of blocks, and thus surface roughness, reveal the very different geologic history of each bodies. On Eros, the surface is dominated by cratering, and impact ejecta controls the distribution of blocks. On Itokawa, the block distribution is a function of the current geopotential of the asteroid and is not associated with individual impact craters. The source of the blocks on Itokawa is likely the catastrophic impact thought to have created Itokawa (Fujiwara et al., 2006), but the current distribution of blocks is due to down-slope movement and thus the current geopotential of Itokawa.

Comparison of the deviograms for Itokawa and Eros (Fig. 10) highlights different surface roughness behavior, with the caveats that the deviogram for Itokawa samples a more restricted range of baselines, and the absolute RMS deviation values depend on the distance metric. The deviogram for Eros is a straight line that increases with increasing baseline implying that the surface roughness at the measured baselines is fractal. On Itokawa the deviogram appears to have less steep slopes at the (albeit restricted range of) longer baselines. This behavior is similar to that observed at long baselines (1–2 km) on the Moon and Mercury (Rosenburg et al., 2011; Fa et al., 2016). The Hurst exponent at baselines of 16 – 32 m from the global deviogram from this study is comparable to that observed from the Highlands regions in (Barnouin-Jha et al., 2008b) at these baselines.

We postulate that this difference in deviograms for Eros and Itokawa may result from the lack of ability of Itokawa to support longer-wavelength topography, in contrast to Eros, which is thought to have a more intact, and stronger, interior (see Fig. 11). For example, Eros shows clear crater cavities (Veverka et al., 1999) while on Itokawa crater morphologies, where seen, are muted (Hirata et al., 2009). In addition, the distribution of Hurst exponents (Fig. 12) is quite different for both asteroids, with Eros having larger Hurst exponents on average than Itokawa. Future missions to rubble-pile asteroids will allow us to explore this hypothesis with new datasets for different asteroids. If the deviograms of rubble-piles and fractured monoliths are different, we may be able to use surface roughness to probe the sub-structure of asteroids. This would be particularly important for asteroids for which mass estimates are poorly known (such as a flyby asteroid mission) but for which have topography data (e.g. from SPC) might be available and sufficient to calculate surface roughness over a range of baselines.

## 5. Conclusions

We measured the global surface roughness of Itokawa from baselines of 8–32 m using topography from the Hayabusa LiDAR, and provide the first global surface roughness maps at the meter-scale for a small-rubble-pile asteroid. A major challenge for establishing surface roughness on small, irregularly-shaped bodies is the choice of distance metric and overcoming this permitted us to use the entire Hayabusa LIDAR dataset for the first time. We investigated several such metrics and our results support the use of the simple Euclidean distance between pairs of points (referred to in this paper as the 'radial-distance' metric). A consequence is that surface roughness can be reliably established only at baselines below those at which the surface curvature of the asteroid becomes important.

We found that the surface roughness of Itokawa is clearly related to the surface geology. The global surface roughness maps at baselines of 8–32 m shows that the highlands on Itokawa have higher surface roughness values than the lowlands. In addition, no significant differences in the surface roughness properties of the head and body of Itokawa were found. However, surface roughness is largest at the equator and smallest at the poles, and also exhibits substantial short-wavelength longitudinal variations. These geographic variations in surface roughness echo those observed in the distribution of boulders on the asteroid, specifically the variation in boulder density with latitude and the lack of differences between the head and the body, reported by Mazrouei et al. (2014). The geographical variations in surface roughness have several implications. First, the similar properties of the head and body in both block density and surface roughness indicate that if the head and body evolved separately and joined at a later date, the current surface topography at scales of 8–32 m is a record of the surface after this join. Second, the differences in surface roughness between the highlands and lowlands indicate estimates of regolith thickness in the lowlands of 1 m to a few meters consistent with previous studies based on a few LIDAR tracks (Barnouin-Jha et al., 2008b). Third, the latitudinal variations in surface roughness may provide insights into the origin of Itokawa, as suggested from similar block count distributions (Mazrouei et al., 2014). Fourth, longitudinal variations in surface roughness, if they continue down to the centimeter-scale, could affect the thermophysical properties of Itokawa and hence influence YORP. This is particularly important as it may help explain mismatch between observed and predicted YORP.

Although the surface roughness of Itokawa and the much-larger fractured-monolith asteroid Eros (Susorney and Barnouin, 2018) both correlate with spatial block density, the distribution of blocks and thus surface roughness are a function of the different geology of rubble-piles and fractured monoliths. On Eros, the block distribution/surface roughness is a function of large impact craters (Thomas, P. C. and Robinson, 2005), while on Itokawa the block distribution/surface roughness is a function of elevation. The deviogram and distributions of Hurst exponents on both bodies are also different, with Eros displaying a more self-affine-like deviogram, while the deviogram for Itokawa appears to shallow in slope at 20–30 m. This difference may result from Itokawa being unable to support long wavelength topography. A wider range of baselines, not available from current topography data for Itokawa, is needed to fully test this observation, but deviograms from upcoming spacecraft encounters at 162173 Ryugu (Tsuda et al., 2013) and 101955 Bennu (Lauretta et al., 2015) will provide insights from other small asteroids.

## Acknowledgments

We would like to acknowledge the helpful reviews by two anonymous reviewers that improved and strengthened the paper. CLJ, HCMS, MGD, and JAS acknowledge OSIRIS-REx Laser Altimeter science support form the Canadian Space Agency. This material is based upon work supported by NASA under Contract NNM10AA11C issued through the New Frontiers Program.

Gaskell, R., Saito, J., Ishiguro, M., Kubota, T., Hashimoto, T., Hirata, N., Abe, S., Barnouin-Jha, O., Scheeres, D., Sep. 2008. Gaskell Itokawa Shape Model V1.0. NASA Planetary Data System 92.

Hirata, N., Barnouin-Jha, O. S., Honda, C., Nakamura, R., Miyamoto, H., Sasaki, S., Demura, H., Nakamura, A. M., Michikami, T., Gaskell, R. W., Saito, J., Apr. 2009. A survey of possible impact structures on 25143 Itokawa. Icarus 200 (2), 486–502.

Kreslavsky, M. A., Head, J. W., Neumann, G. A., 2014. Kilometer-scale topographic roughness of Mercury: Correlation with geologic features and units. Geophysical Research Letters.

Kreslavsky, M. A., Head, J. W., Neumann, G. A., Rosenburg, M. A., Aharonson, O., Smith, D. E., Zuber, M. T., Oct. 2013. Lunar topographic roughness maps from Lunar Orbiter Laser Altimeter (LOLA) data: Scale dependence and correlation with geologic features and units. Icarus 226 (1), 52–66.

Lauretta, D. S., Bartels, A. E., Barucci, M. A., Bierhaus, E. B., Binzel, R. P., Bottke, W. F., Campins, H., Chesley, S. R., Clark, B. C., Clark, B. E., Cloutis, E. A., Connolly, H. C., Crombie, M. K., Delbó, M., Dworkin, J. P., Emery, J. P., Glavin, D. P., Hamilton, V. E., Hergenrother, C. W., Johnson, C. L., Keller, L. P., Michel, P., Nolan, M. C., Sandford, S. A., Scheeres, D. J., Simon, A. A., Sutter, B. M., Vokrouhlický, D., Walsh, K. J., Apr. 2015. The OSIRIS-REx target asteroid (101955) Bennu: Constraints on its physical, geological, and dynamical nature from astronomical observations. Meteoritics & Planetary Science 50 (4), 834–849.
23

**Figures**

Table 1: Correlation coefficients for all correlations with the surface roughness of Itokawa at the baselines investigated in this study.

| Baseline | Boulder Spatial Density | Topography | Slope | Geopotential |
|---|---|---|---|---|
| 8 m  | 0.23 | 0.43 | 0.46 | -0.44 |
| 16 m | 0.24 | 0.48 | 0.66 | -0.60 |
| 24 m | 0.10 | 0.50 | 0.69 | -0.69 |
| 32 m | 0.05 | 0.47 | 0.66 | -0.69 |

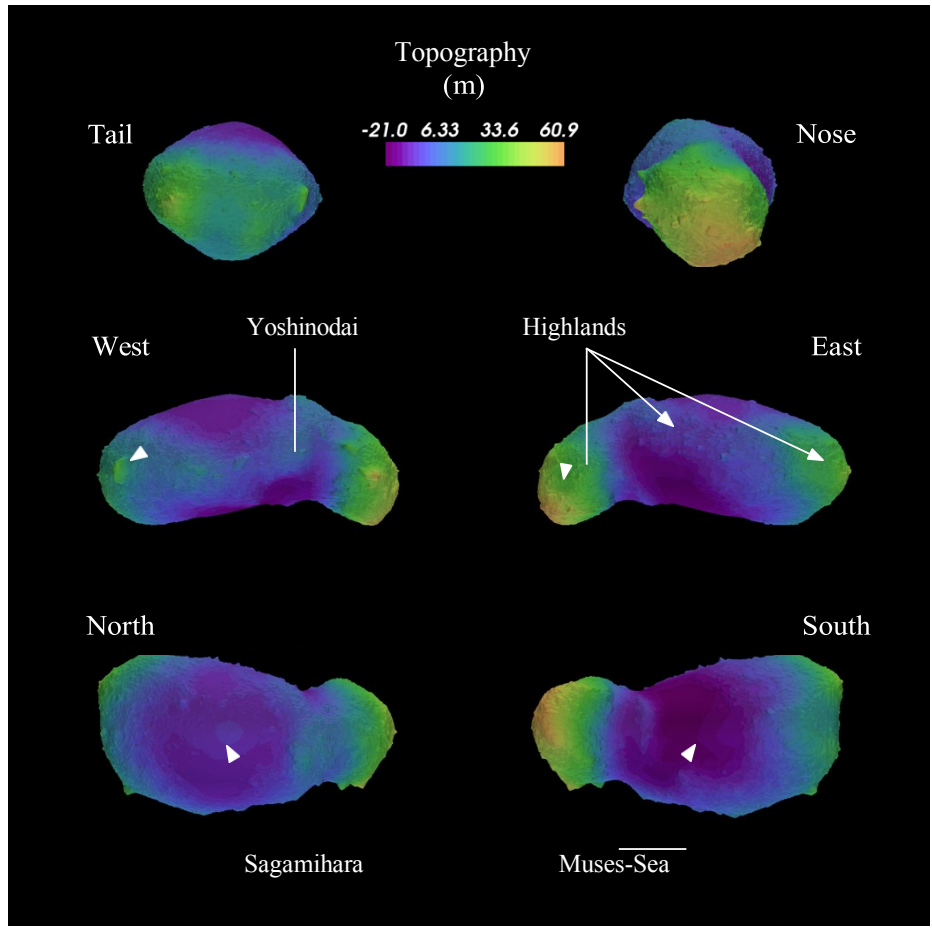

Figure 1: The topography of Itokawa shown on a 49,152-plate shape model (Gaskell et al., 2008). The surface of Itokawa can be broadly separated into lowlands (Muses–Sea and Sagamihara) and highlands. The largest boulder on the surface, Yoshinodai is labeled and is 50 × 30 × 20 m (Saito et al., 2006).

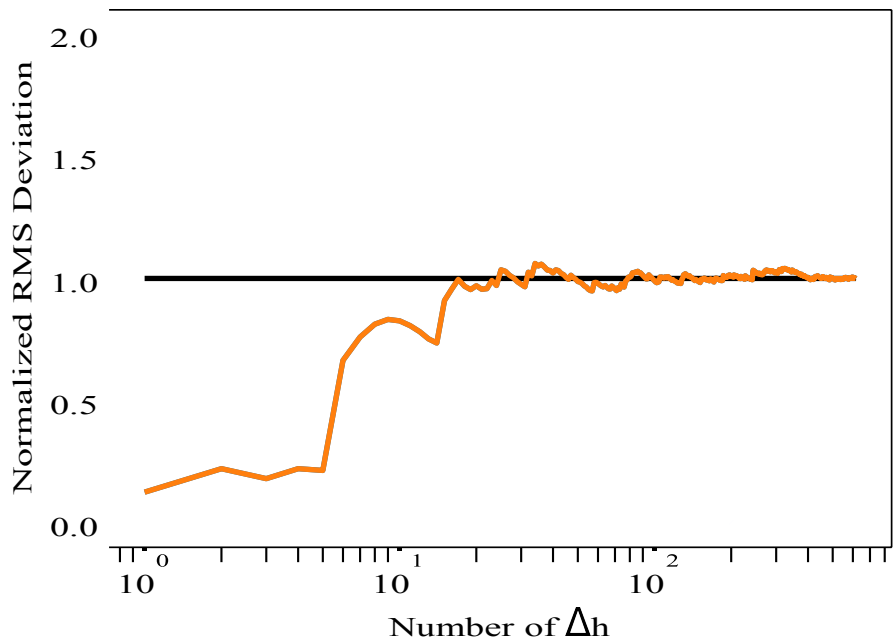

Figure 2: The stability of the RMS deviation estimate (normalized to the final value) as a function of the number of estimates, $n$ of $\Delta h$, at a random location on the surface of Itokawa. The stability of the RMS deviation was investigated at several locations across Itokawa and results indicate that it converges at about $n = 200$.

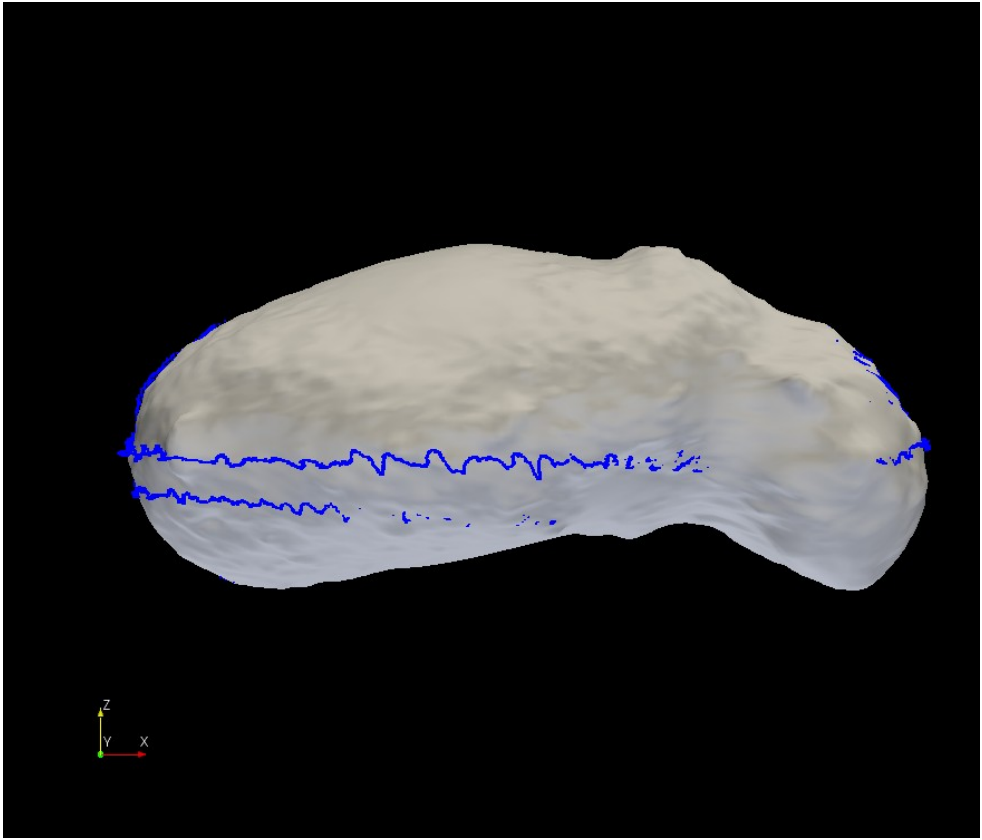

Figure 3: An example of the 'wandering' Hayabusa LIDAR tracks. This is track *cdr_f 2005_10_01* from the PDS small body node (*https : //sbnarchive.psi.edu/pds3/hayabusa/HAY_A_LIDAR_3_HAY_LIDAR_V_2_0/data/cdr/*)

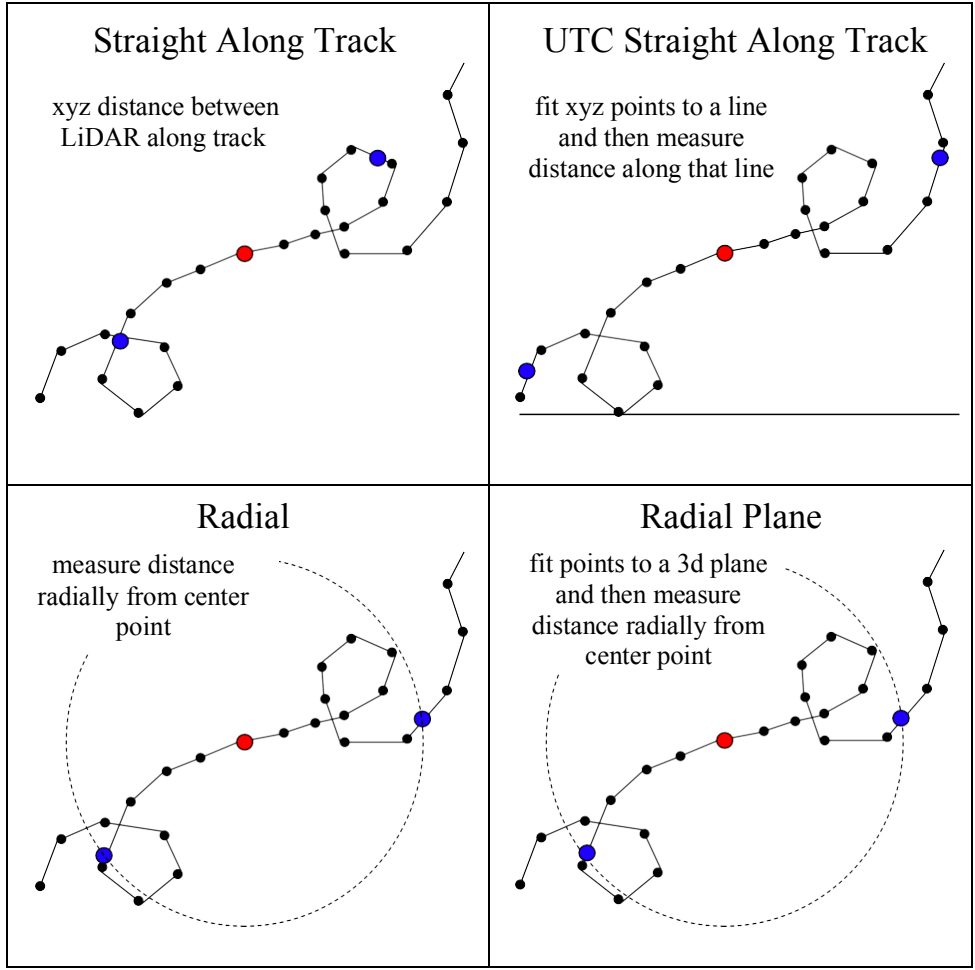

Figure 4: Schematic of the four different distance estimates used in this paper. The black points are the LIDAR returns, the red point is the point of interest, the two blue points are the points over which $\Delta h$ was computed, and the dashed black circles in the lower two schematics represent one baseline away from the point of interest.

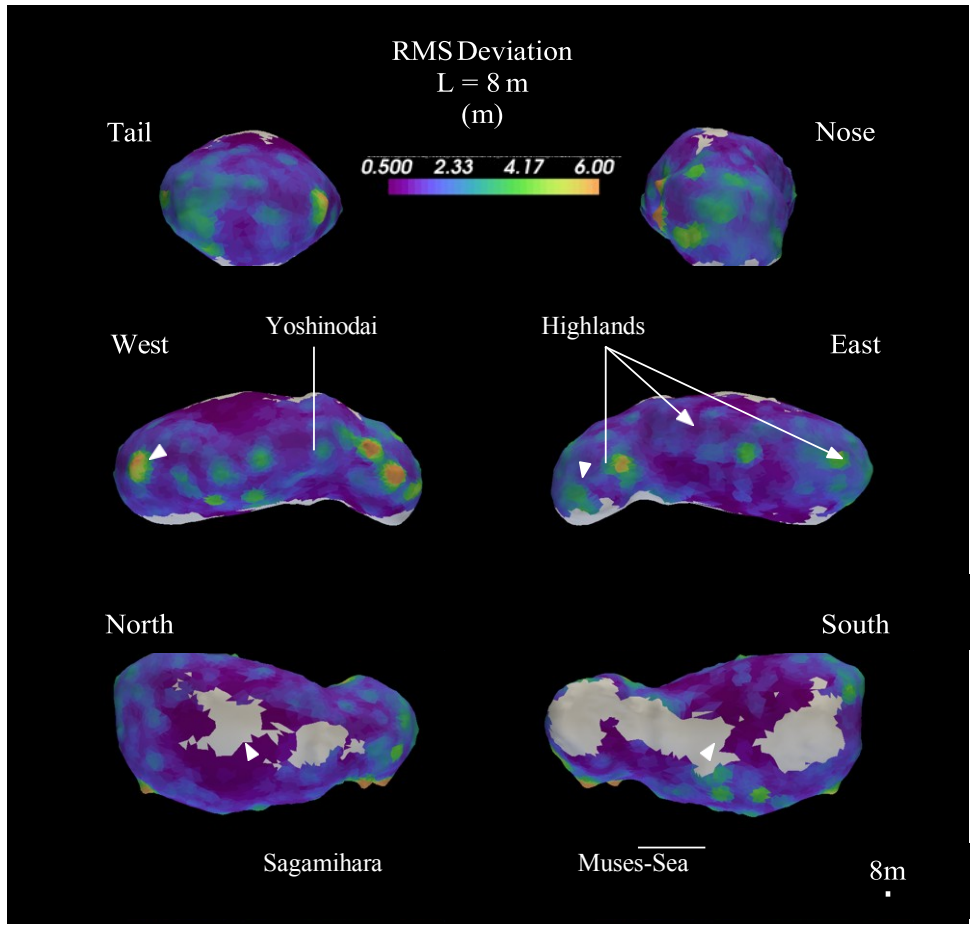

Figure 5: The surface roughness at a baseline of 8 m. Gray indicates that no estimate was available due to low track density. Note the relatively high surface roughness in the highlands and the relatively low surface roughness in the Muses–Sea and Sagamihara (where there is adequate $\Delta h$ to calculate surface roughness). The surface roughness is elevated associated with boulders such as Yoshinodai.

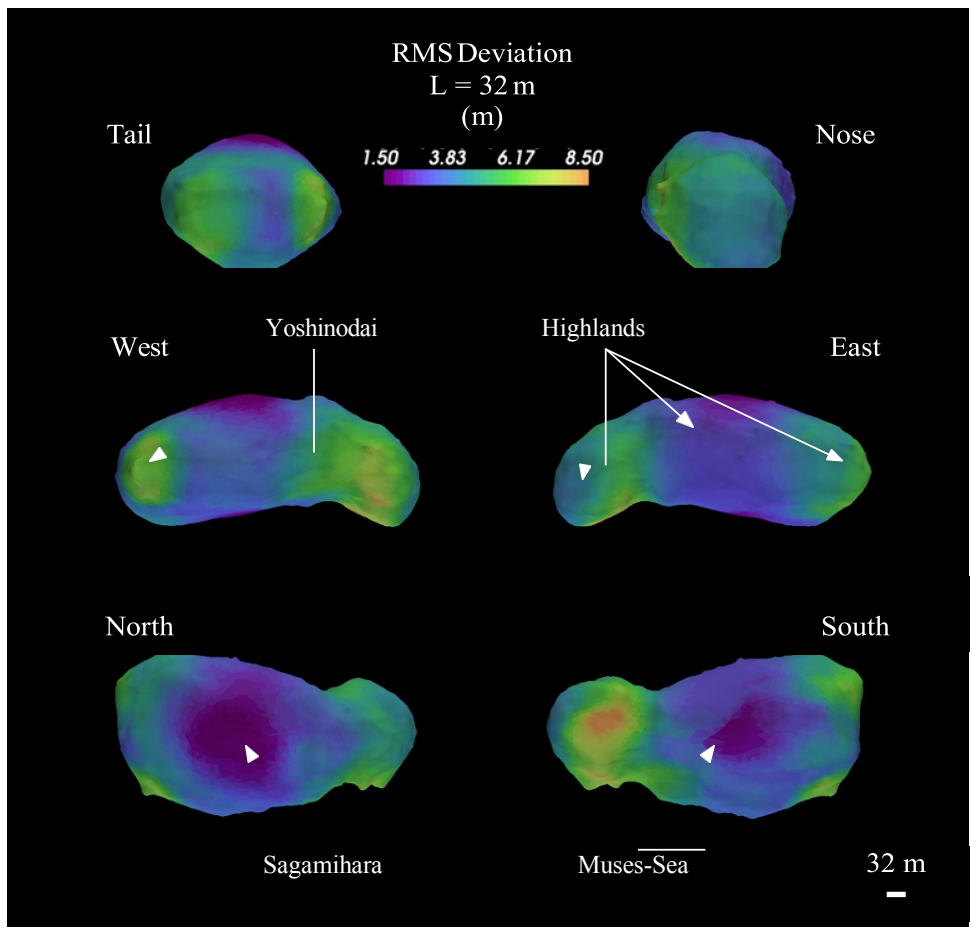

Figure 6: The surface roughness at a baseline of 32 m. The surface roughness follows the same pattern as the 8 m baseline with the highlands and lowlands having a bimodal surface roughness relationship. Individual boulders at the 32 m baseline are less clear and instead are associated with the overall higher surface roughness of the highlands.

Figure 7: The Hurst exponent for baselines from 8 m to 32 m. Note that the Hurst exponent is very low around boulders and varies across the asteroid surface.

Figure 8: A map of the block spatial density from Mazrouei et al. (2014). The blocks follow the highland/lowland dichotomy.

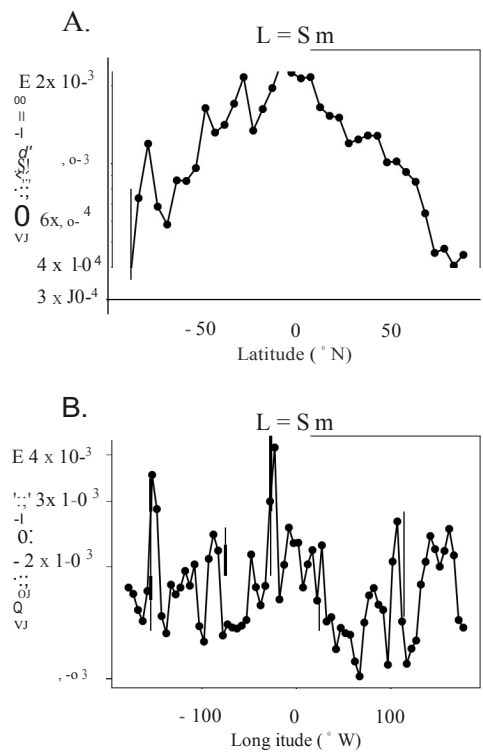

Figure 9: The RMS deviation calculated in 5° degree bins as a function of (A) the latitude and (B) longitude of Itokawa at a baseline of 8 m. Surface roughness is highest at the equator and shows substantial variation with longitude but no clear pattern.

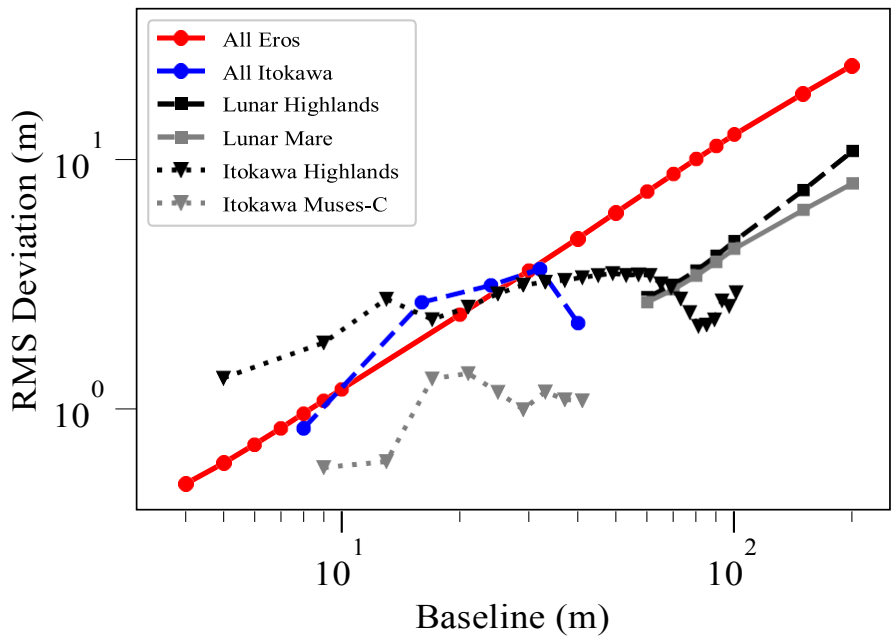

Figure 10: Deviograms for Itokawa, Eros, and the Moon. The All Eros, the Lunar Highlands, and the Lunar Mare data are from Susorney and Barnouin (2018). The Itokawa Muses–Sea and Itokawa Highlands are from Barnouin-Jha et al. (2008b).

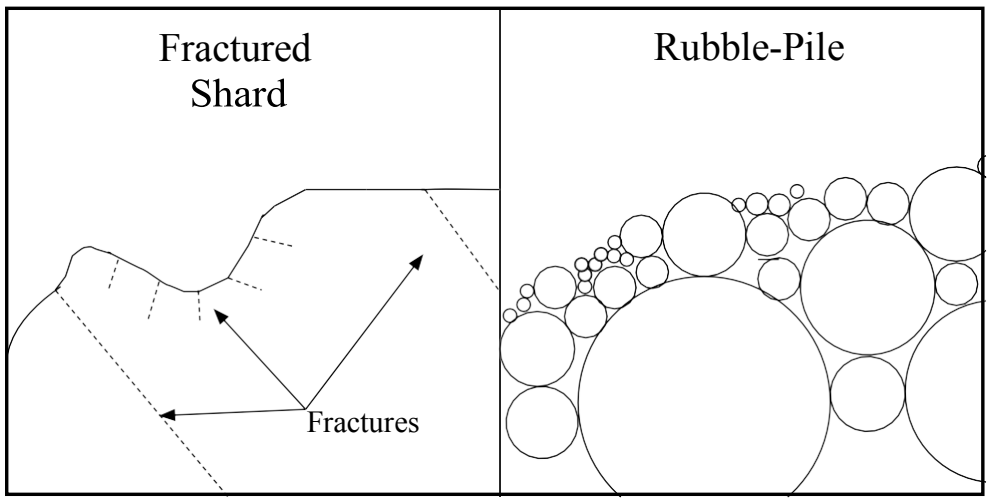

Figure 11: A schematic of the proposed near-surface structure of Eros and Itokawa. The fractured shard (Eros) is better able to support topography, in particular topography due to impact craters, while the rubble-pile (Itokawa) supports a smaller range of topography and impact craters do not have the same topographic expression (Hirata et al., 2009).

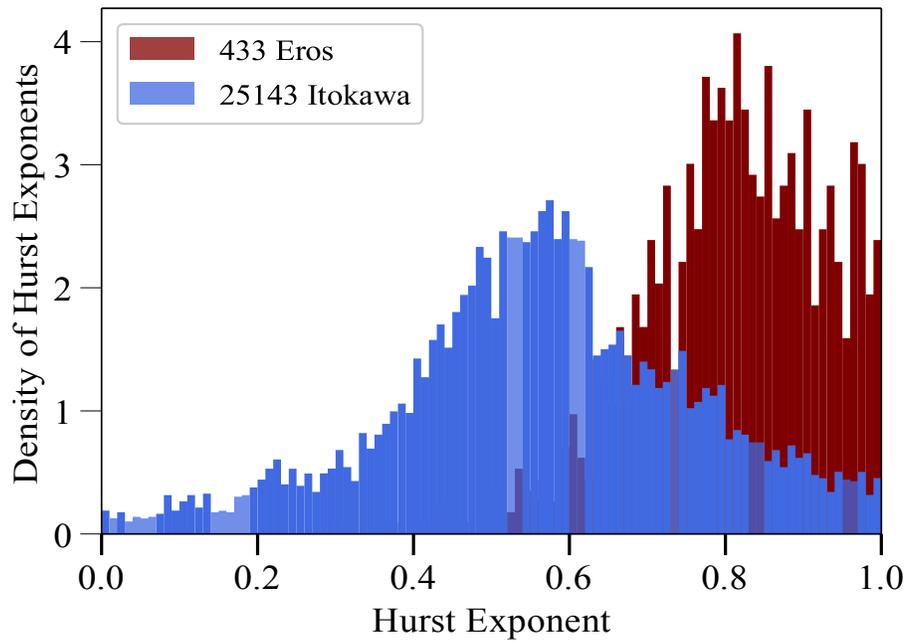

Figure 12: Histograms of the distribution of Hurst exponents on Itokawa and 433 Eros (Susorney and Barnouin, 2018) from the Hurst exponent maps. Note how different the peaks and spread in Hurst exponents are for the two asteroids.